\documentclass{emulateapj}
\usepackage{apjfonts}
\usepackage{graphicx}
\usepackage{subfigure}
\newcommand{\secpoint}{\mbox{$''\mskip-7.6mu.\,$}}

\slugcomment{Draft Version July 20, 2013}

\begin{document}

\title{SALT Long-slit Spectroscopy of Luminous Obscured Quasars: An Upper Limit on the Size of the Narrow-Line Region?}

\shorttitle{SALT Spectroscopy of Obscured Quasars}
\shortauthors{HAINLINE ET AL.}

\author{\sc Kevin N. Hainline, Ryan C. Hickox}
\affil{Department of Physics and Astronomy, Dartmouth College, Hanover, NH 03755}

\author{\sc Jenny E. Greene}
\affil{Department of Astrophysical Sciences, Princeton University, Princeton, NJ 08544}

\author{\sc Adam D. Myers}
\affil{Department of Physics and Astronomy, University of Wyoming, Laramie, WY 82071}

\author{\sc Nadia L. Zakamska}
\affil{Department of Physics and Astronomy, The Johns Hopkins University, Baltimore, MD 21218}

\author{}

\begin{abstract}

We present spatially resolved long-slit spectroscopy from the Southern African Large Telescope (SALT) to examine the spatial extent of the narrow-line regions (NLRs) of a sample of 8 luminous obscured quasars at $0.10 < z < 0.43$. Our results are consistent with an observed shallow slope in the relationship between NLR size and $L_{[\mathrm{OIII}]}$, which has been interpreted to indicate that NLR size is limited by the density and ionization state of the NLR gas rather than the availability of ionizing photons. We also explore how the NLR size scales with a more direct measure of instantaneous AGN power using mid-IR photometry from WISE, which probes warm to hot dust near the central black hole and so, unlike [OIII], does not depend on the properties of the NLR. Using our results as well as samples from the literature, we obtain a power-law relationship between NLR size and $L_{8\mu \mathrm{m}}$ that is significantly steeper than that observed for NLR size and $L_{[\mathrm{OIII}]}$.  We find that the size of the NLR goes approximately as $L^{1/2}_{8\mu\mathrm{m}}$, as expected from the simple scenario of constant-density clouds illuminated by a central ionizing source. We further see tentative evidence for a flattening of the relationship between NLR size and $L_{8 \mu\mathrm{m}}$ at the high luminosity end, and propose that we are seeing a limiting NLR size of $10 - 20$ kpc, beyond which the availability of gas to ionize becomes too low. We find that $L_{[\mathrm{OIII}]} \sim L_{8 \mu\mathrm{m}}^{1.4}$, consistent with a picture in which the $L_{[\mathrm{OIII}]}$ is dependent on the volume of the NLR. These results indicate that high-luminosity quasars have a strong effect in ionizing the available gas in a galaxy. 

\end{abstract}

\keywords{galaxies: active -- quasars: emission lines}

\section{Introduction}
\label{sec:intro}

	The size of the extended emission-line regions around active galactic nuclei (AGNs) has been the focus of much recent research in an attempt to understand the effect that a central supermassive black hole can have on the gas on galaxy-wide scales. The hard ionizing radiation from an accreting black hole can affect galactic gas at large distances from the central source, producing distinct emission features that can be carefully studied with imaging and spectroscopy. Regions of ionized gas that extend out to tens of kpc were initially discovered around radio-loud quasars by \citet{wampler1975} and \citet{stockton1976}. An extensive campaign by \citet{stocktonmackenty1987} utilized narrow-band imaging of nearby luminous quasars to find that ionized emission around radio-loud quasars extended out to between 10 - 50 kpc in a large fraction of the objects. \citet{bennert2002} used Hubble Space Telescope (HST) imaging for a sample of radio-quiet quasars to detect ionized (as traced by [OIII]$\lambda$5007 emission) gas out to 10 kpc, and also first described a relationship between the size of the narrow-line region and $L_{[\mathrm{OIII}]}$, which was later confirmed by \citet{schmitt2003} using HST imaging of a larger sample of Seyfert 1 and Seyfert 2 galaxies at lower luminosities. This relationship, which is similar to one found between broad-line region size and AGN continuum luminosity in (unobscured) Type I AGNs, provides insight into the properties of the NLR gas such as the covering fraction or density.

	This extended ionized emission is often identified as originating from the narrow-line region (NLR) around the AGN. The NLR was first discovered using narrow-band imaging, but most recent studies of the size and kinematics of the narrow-line region rely on spectroscopy. While imaging allows for a detailed study of the NLR morphology, the broad filters used make understanding the true size of the NLR difficult, and spectroscopy can probe specific emission features that trace the NLR directly. Long slit spectroscopy can also serve as a probe of the physical properties and kinematics of the NLR gas. Both \citet{fraquelli2003} and \citet{bennert2006} used long-slit spectroscopy to examine samples of nearby Seyfert galaxies, and explored the size of the NLR as well as trends of ionization parameter and electron density with spatial position. 
	
	It is important, however, to understand whether there exists a limit to the NLR size in powerful AGNs. The slope predicted by \citet{bennert2002} and \citet{schmitt2003} for the size-luminosity relationship predicts unphysically large NLR sizes at high luminosity \citep{netzer2004}. In the highest luminosity sources discussed by \citeauthor{netzer2004}, the NLR size would be $R_{NLR} = 20 - 100$\,kpc, which would exceed the sizes of all but the largest cD galaxies. To explore the true extent of the largest NLRs, it is important to examine the most powerful quasars. Large NLRs have been observed in individual Type I quasars, such as MR 2251-178 \citep[$z = 0.0640$,][]{kreimeyer2013}, but studies of larger samples of Type I quasars are often challenging due to contamination by emission from the central source \citep{husemann2013}. In Type II quasars, an obscuring medium is thought to block the UV/optical emission from near the black hole along our line of sight, making these objects ideal targets for studying the extent and kinematics of the NLR. Due to their rarity, studies of Type II quasars have mainly focused on individual objects: \citet{villarmartin2010} targeted the Type II quasar SDSS J0123+00 ($z = 0.399$) using long-slit spectroscopy, and found that this quasar had ionized emission out to $\sim 133$ kpc, although this galaxy is in an interacting system. Fortunately, recent results from the Sloan Digital Sky Survey \citep[SDSS,][]{york2000} have made possible the construction of a large sample of low redshift ($z < 0.8$) obscured quasars selected based on emission line ratios and $L_{[\mathrm{OIII}]}$\citep{zakamska2003,reyes2008}. The full sample, which encompasses 1000 galaxies, has been studied with a variety of instruments \citep[see references in][]{liu2013}. 
	
	Initial long-slit spectroscopy for members of the Type II quasar sample was performed on a selection of mostly radio-quiet low-redshift ($z < 0.4$) obscured quasars in \citet{greene2011}, who sought to explore both the spatial extent and the resolved kinematics of the NLR gas. These authors examined the spatial extent of the [OIII]$\lambda$5007 line across each quasar in their sample and observed ionized gas at large radii, and then used the results for high luminosity quasars to obtain a much shallower slope in the relationship between NLR-size and L$_{[\mathrm{OIII}]}$ than was observed by \citet{bennert2006}. The authors attributed this shallow slope to the NLR gas being matter-bounded, where the NLR size is limited by the density of the gas instead of the availability of ionizing photons.
	
	These results were confirmed for a similar, but more luminous, sample of radio-quiet obscured quasars by \citet{liu2013} using integral field unit (IFU) spectroscopy. These authors measured [OIII] emission out to an average of $14 \pm 4$ kpc from the nucleus. Based on the observed sizes of nebulae, the flat index of the size-luminosity relationship ($R\propto L_{[\mathrm{OIII}]}^{0.25}$) and the decline of the [OIII]/H$\beta$ in the outer parts of the nebulae, these authors propose a model where the NLR clouds transition to a low-density, matter-bounded region beyond $\sim7$ kpc. \citet{husemann2013} used IFU data to carefully analyze a sample of low-redshift radio-quiet Type I quasars and found that the typical size of the extended NLR is around 10 kpc, comparable to the sizes observed in Type II quasars at similar [OIII] luminosities.  
		
	In this paper, we examine a sample of Type II radio-quiet quasars from the \citet{zakamska2003} and \citet{reyes2008} samples using long-slit spectroscopy with the Southern African Large Telescope (SALT), following the analysis presented in \citet{greene2011}. We examine the spatial extent of the NLR, and explore how the size of the NLR varies as a function of mid-IR luminosity, which is a more direct measure of intrinsic AGN power. While the [OIII] emission line is the most common optical tracer of AGN strength, it can suffer from strong dust obscuration, and more importantly $L_{[\mathrm{OIII}]}$ is dependent on the properties of the NLR, and is limited by the quantity, density, and ionization state of the gas present in a galaxy. We can use data in the mid-IR to directly probe AGN luminosity and thus the amount of ionizing photons that exist for a given NLR. Recently, the Wide Field Infrared Explorer \citep[WISE,][]{wright2010} has provided photometry across the entire sky at four infrared wavebands: 3.4, 4.6, 12, and 22 $\mu$m. In powerful AGN, emission at these mid-IR wavelengths traces warm to hot dust emission very near the central engine \citep{pier1993}, and IR emission has been shown to correlate strongly with AGN soft X-ray emission \citep[e.g.][]{krabbe2001, lutz2004, horst2008, asmus2011, matsuta2012}. The IR acts as an excellent proxy for AGN power that does not depend on NLR properties. In this paper we use WISE all-sky coverage to estimate IR luminosities for a large sample of spectroscopically-confirmed AGNs from the literature to explore how the size of the NLR varies as a function of this direct measure of AGN luminosity.
	
	We describe our sample in Section \ref{sec:sample}, discuss the spatial sizes of the observed NLRs in Section \ref{sec:nlrsizes}, and then discuss our results and draw conclusions in Section \ref{sec:discussion}. Throughout, we assume a standard $\Lambda$CDM cosmological model with $H_0 = 70$ km s$^{-1}$ Mpc$^{-1}$, $\Omega_{M} = 0.3$, and $\Omega_{\Lambda} = 0.7$.
	
\section{Quasar Sample, Observations and Data Reduction}
\label{sec:sample}

\begin{deluxetable*}{lccccccc}
\tabletypesize{\scriptsize}
\tablecaption{Sample and Observations \label{tab:sample}}
\tablewidth{0pt}
\tablehead{
\colhead{SDSS Name} & \colhead{$z$} & \colhead{log($L_{[\mathrm{OIII}]}$)} & \colhead{log($L_{8\mu m}$)} & \colhead{Obs. Date}
& \colhead{seeing} & \colhead{t$_{s}$} & \colhead{P.A.\tablenotemark{a}} \\
\colhead{} & \colhead{} & \colhead{(ergs/s)} & \colhead{(erg/s)} & \colhead{} & \colhead{($''$)} & \colhead{(s)} & \colhead{}   
}
\startdata

J024940.2--082804.6 & 0.268 & 42.5 & 44.7 & 2011 Nov 22 & 1.9 & 2964 & 97 \\
J031428.3--072517.8 & 0.208 & 42.7 & 44.4 & 2011 Nov 29 & 2.3 & 2566 & 235 \\
J033606.7--000754.8 & 0.432 & 42.3 & 44.9 & 2012 Jan 25 & 0.9 & 2400 & 140 \\
J081125.8+073235.4 & 0.350 & 42.4 & 44.4 & 2011 Dec 31 & 2.1 & 1800, 2100 & 130, 220\tablenotemark{b} \\
J084107.1+033441.3 & 0.274 & 42.3 & 43.9 & 2011 Dec 29 & 1.8 & 2400, 2100 & 230, 140\tablenotemark{b} \\ 
J084135.1+010156.3 & 0.111 & 42.4 & 44.5 & 2012 Apr 16 & 1.4 & 1500 & 235 \\
J110012.4+084616.4 & 0.100 & 42.7 & 45.0 & 2012 Apr 17 & 2.1 & 2400 & 150 \\
J122217.9--000743.8 & 0.173 & 42.9 & 45.0 & 2012 Apr 17 & 2.1 & 1170 & 250 

\enddata
\tablenotetext{a}{Position Angle, in degrees east of north.}
\tablenotetext{b}{This object was observed at two position angles for comparing the sizes measured along different axes.}

\end{deluxetable*}

	Our sample consists of 8 quasars. We start from a set of Type II quasars selected using SDSS spectroscopy by \citet{zakamska2003} and \citet{reyes2008}. We first selected candidates with log $L_{[\mathrm{OIII}]}  > 42.25$ erg s$^{-1}$ to maximize the likelihood of resolving the NLR, and at $0.2 < z < 0.45$ such that [OIII] would fall into our spectroscopic coverage. To ensure maximum visibility with SALT, we then limited our candidates to those with RA $< 150^{\circ}$ and DEC $< +10^{\circ}$. We used observations at 1.4 GHz from the FIRST survey \citep{becker1995, white1997}, to select only those objects which are radio-quiet as determined by their position on the $L_{[\mathrm{OIII}]}$ vs $\nu$L$_{\nu}$(1.4 GHz) diagram \citep{xu1999, zakamska2004}. Finally, we chose targets that have a range in the WISE 12$\mu$m / $L_{[\mathrm{OIII}]}$ ratio to examine how NLR properties vary as a function of IR luminosity. The sample and observations are given in Table \ref{tab:sample}, and throughout this paper we will refer to these objects using shortened names. We also selected two objects (J0841+0101 and J1222--0007) that were previously observed by \citet{greene2011} for comparison to our observations and methodology.

	The quasar sample was observed with the Robert Stobie Spectrograph \citep[RSS;][]{kobulnicky2003,smith2006} on SALT in Sutherland, South Africa. The observations were carried out in long-slit mode in queued campaigns between November of 2011 and April of 2012. For these observations, we used a 1\secpoint25 slit and the RSS PG1300 grating, which provides a spectral resolution of 3.2\,\AA\ at 6000\,\AA. The seeing was measured using stars observed in the acquisition images taken just before the observations, and was typically $\sim 2''$ for the runs. The objects were observed between 20 and 50 minutes each, and two of the targets were observed with two slit position angles (P.A.). The slit positions were chosen with respect to the original SDSS imaging to correspond to the observed major and minor axes when the galaxy was resolved, or, in some instances, the P.A. was chosen to align the slit with nearby galaxies to determine if they were companions. Unfortunately, due to the fixed nature of the primary mirror at the SALT telescope and the strong variation in effective aperture with time and source position, absolute flux calibration is not possible using the SALT data alone. We perform an approximate flux calibration on our spatial plots in order to compare the spatial extent of the objects as described in \S \ref{sec:measurements}. 

	Data reduction was performed using standard IRAF scripts\footnote{IRAF is distributed by the National Optical Astronomy Observatory, which is operated by the Association of Universities for Research in Astronomy (AURA) under cooperative agreement with the National Science Foundation.}. The data were gain-corrected, bias-corrected, and mosaiced with the SALT reduction pipeline. We flat-fielded the data, applied a wavelength solution using arc lamp spectra, and used the multiple images made for each target for cosmic ray removal. Finally, the two-dimensional spectra were background subtracted and combined using a median combine. 
		
\section{Narrow-Line Region Sizes}
\label{sec:nlrsizes}

\subsection{Size Measurements}
\label{sec:measurements}

\begin{figure}[htbp]
	\epsscale{1.2} 
	\plotone{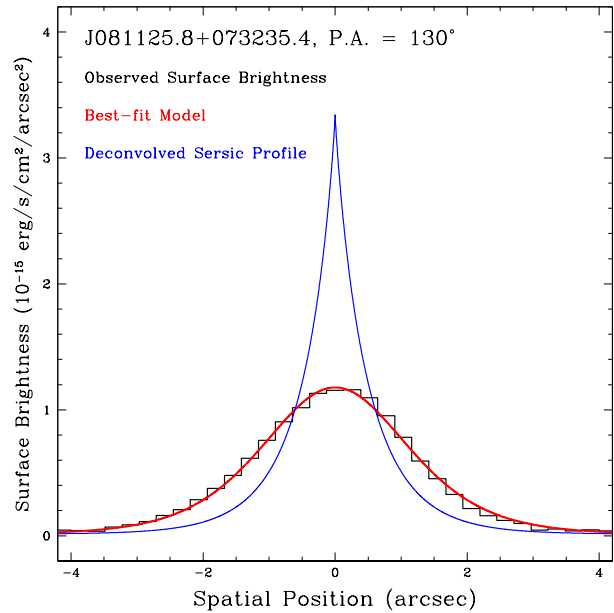} 
	\caption{Observed surface brightness profile for J0811+0732 (in black), plotted with the best-fit model (in red), and the deconvolved intrinsic profile (in blue). The seeing was modeled by a Moffat fit, and the intrinsic model was represented by a S\'{e}rsic profile. 
	\label{fig:deconvolution}} 
	\epsscale{1.}
         \end{figure}

	We can use the high S/N spectra to measure the physical extent of the narrow-line region in these quasars, as a way of understanding the impact the AGN is having on its host galaxy. We started by following the prescription of \citet{greene2011} and created a spatial profile for the [OIII]$\lambda$5007 feature by collapsing the 2D spectrum in the wavelength direction with a width twice the FWHM of [OIII] in order to increase the S/N of the emission on the outer edges of the galaxy. Each of our objects had an observed [OIII] spatial extent that was larger than our measured seeing for the observation. To account for seeing effects, we sought to fit each observed profile using a model for the true spatial profile convolved with a model of the measured seeing. We used multiple models to describe both the true profile and the seeing in order to see how choice of model affected the resulting NLR size. For the spatial profiles we used both a S\'{e}rsic profile and a Voigt profile. Similarly, we examined two different models for the seeing: a Gaussian profile and a Moffat profile. While the final fits had similar reduced $\chi^2$ values, the resulting sizes varied by an average of 20\% depending on the models used. These size differences are a result of the extended wings in the observed profiles resulting from the seeing (when modeled by a Moffat profile) rather than being a feature of the true spatial profile (when modeled by a Voigt profile). For our analysis, we chose to use a S\'{e}rsic profile to model the spatial extent of the [OIII] emission convolved with a Moffat profile for the seeing, which has been shown to robustly model point-spread functions \citep{trujillo2001}. To estimate the Moffat seeing parameters, we fit Moffat profiles to the stars in the acquisition images, and took an average of the Moffat $\alpha$ and $\beta$ values for use in the convolution. The SALT point-spread function measured from the acquisition images is similar between the different observations, and is well characterized with a Moffat fit. We plot a representative surface brightness profile for the quasar J0811+0732, along with both the best-fit model and the deconvolved S\'{e}rsic profile for this object in Figure \ref{fig:deconvolution}. This figure demonstrates the importance of accounting for the seeing when measuring the intrinsic size of the galaxies in our sample.

	In order to compare the extent of the [OIII] emission line to those measured by other authors, we sought to measure the line extent using a method that was not dependent on the depth of our observation, unlike some prescriptions for NLR size for which the width is defined using a specific flux measured above the noise. We chose the parameter described in \citet{liu2013}, $R_{\mathrm{int}}$, which is calculated as the size of the galaxy at a limiting surface brightness corrected for cosmological dimming of $10^{-15}$ erg s$^{-1}$ cm$^{-2}$ arcsec$^{-2}$. In order to calculate $R_{\mathrm{int}}$, we needed to flux calibrate our spatial profiles. 

	We performed flux calibration by comparing our measured spectra to the SDSS spectra. The SDSS spectra are measured using $3''$ diameter fibers, while our spectra were measured using a 1\secpoint25 slit, so we have to correct for this aperture difference as well as for differences in seeing. For each SDSS spectrum, we measured the flux in the [OIII]$\lambda$5007 line using the same width (twice the FWHM) that was used to collapse our SALT spectra. With this total flux, we made the assumption that each object was circularly symmetric on the sky, and that the S\'{e}rsic profile parameters we measured (deconvolved with the seeing) characterized the spatial extent of the true profile for each of our objects. Our assumption of circular symmetry on the sky is supported by the results from \citet{liu2013}, who found that the majority of the ionized gas nebulae from their sub-sample of radio-quiet quasars drawn from the \citet{reyes2008} sample are close to circularly symmetric. We also find similar size measurements made for the objects in our sample which were observed with two position angles. We use the best-fit S\'{e}rsic profile from our modeling of the SALT data to convolve each object by the typical seeing for SDSS spectroscopy ($\sim2''$), and then estimated the fraction of the flux that was observed using the 3$''$ diameter SDSS fiber on the sky. In a similar way, we also convolved the S\'{e}rsic profile with the SALT seeing for each observation, and calculated the fraction of the flux that was covered by the 1\secpoint25 SALT RSS slit. Using these fractions, we calculated correction factors for our SALT spatial profiles such that the total observed flux represented the SDSS flux measured in the same wavelength range. This process is not possible for the two merging quasars in our sample (J0841+0101 and J1222--0007), as our assumption of circular symmetry on the sky is incorrect. These objects are not included in our plots of $R_{\mathrm{int}}$ versus luminosity. We use the deconvolved, flux calibrated surface brightness profiles for each of the remaining objects to calculate $R_{\mathrm{int}}$, which are given in Table \ref{tab:NLRsize}. 
	
	For each of our objects, we created 500 fake spatial profiles using the uncertainties on the observed SALT and SDSS spectral line fluxes, and for each fake spatial profile we ran our flux calibration and $R_{\mathrm{int}}$ measuring procedure. We observed typical fractional uncertainties of around 0.01, which are unphysically small. We have not accounted for multiple sources of error that would serve to increase our uncertainty. For instance, as we discuss above, by using a Voigt profile along with a Gaussian description of the seeing, the NLR sizes are on average 20\% larger. This size difference must be included in our uncertainties, as our fits with different models had similar reduced $\chi^2$ values, indicating that the parameterization is non-unique. Furthermore, while there is evidence that these galaxies are symmetric, the true shape of the objects cannot be understood from an individual longslit spectrum, and non-symmetric morphologies would significantly affect the observed NLR size. We conservatively chose to adopt a 50\% error on our size measurements, which proves to be similar to the observed dispersion in $R_{\mathrm{int}}$ for objects of a similar $L_{[\mathrm{OIII}]}$ discussed in Section \ref{sec:nlrsizeloiii}.  
		 
\subsection{The Relationship between NLR size and $L_{[\mathrm{OIII}]}$}
\label{sec:nlrsizeloiii}

	\begin{figure*}[htbp]
	\epsscale{0.9} 
	\plotone{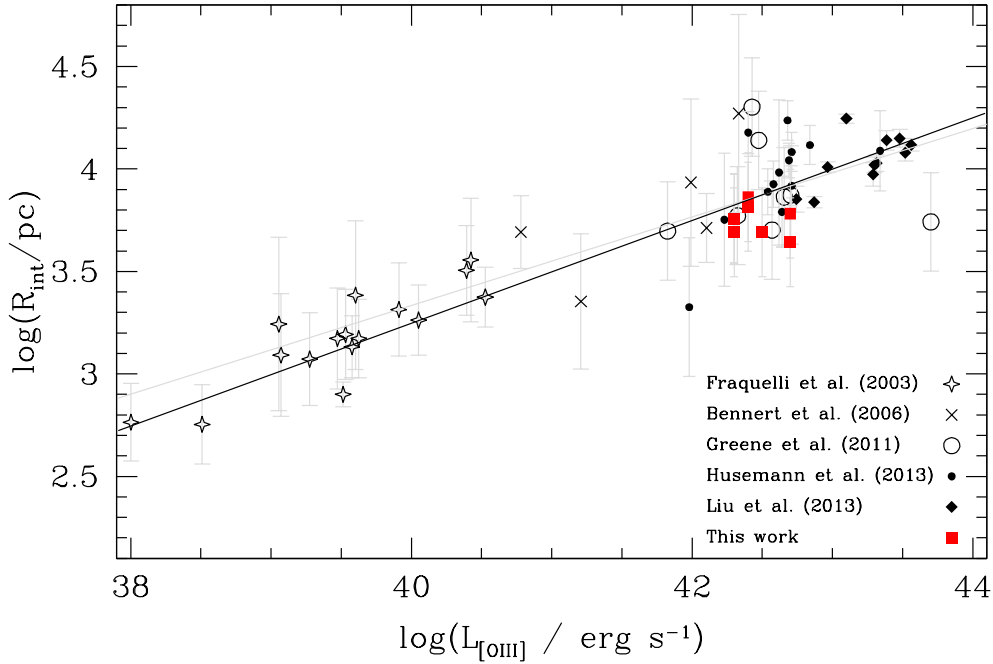} 
	\caption{Radii of the [OIII]$\lambda$5007-emitting region plotted against $L_{[\mathrm{OIII}]}$. The NLR size is represented by $R_{\mathrm{int}}$, which is defined as the size of the object at a limiting surface brightness corrected for cosmological dimming of $10^{-15}/(1+z)^{4}$ erg s$^{-1}$ cm$^{-2}$ arcsec$^{-2}$ \citep[see][]{liu2013}. Our sample of SALT obscured quasars are shown with red squares. We overplot AGN samples from \citet[][nearby Seyfert 2 galaxies, open stars]{fraquelli2003}, \citet[][Seyfert 2 galaxies, Xs]{bennert2006}, obscured quasars, \citep[][open circles]{greene2011}, \citet[][Type I quasars, filled circles]{husemann2013}, and \citet[][obscured quasars, black diamonds]{liu2013}. The best fit from \citet[][slope = $0.22 \pm 0.04$]{greene2011} is shown with a grey line, and the fit from \citet[][slope = $0.25 \pm 0.02$]{liu2013} is shown with a black line. We do not plot merging AGN sources or objects with upper limits on their size measurements. Our SALT sources are consistent with the shallow slope for AGNs between NLR size and $L_{[\mathrm{OIII}]}$ seen in \citet{greene2011} and \citet{liu2013}.
	\label{fig:NLRsize}} 
	\epsscale{1.}
         \end{figure*}

	The size of the NLR is dependent on both the properties of the central ionizing source as well as the distribution, geometry, and kinematics of the host galaxy ISM. We plot the derived NLR sizes against $L_{[\mathrm{OIII}]}$ in Figure \ref{fig:NLRsize}. Also plotted on the figure are AGNs taken from the literature, and in order to make a comparison to galaxies at different redshifts, we follow the prescription of \citet{liu2013} and correct for cosmological dimming by plotting $R_{\mathrm{int}}$ = $10^{-15}/(1+z)^{4}$ erg s$^{-1}$ cm$^{-2}$ arcsec$^{-2}$. We include $R_{\mathrm{int}}$ values for obscured quasars measured using long-slit spectroscopy from \citet[][$z \sim 0.1 - 0.4$]{greene2011}, as well as values for obscured radio-quiet quasars measured using IFU observations from \citet[][$z \sim 0.3 - 0.6$]{liu2013}. We have also plotted the $R_{\mathrm{int}}$ measurements from long-slit spectroscopy for two samples of local Seyfert 2 galaxies from both \citet{fraquelli2003} and \citet{bennert2006}. For these samples, the value of $R_{\mathrm{int}}$ has been remeasured from the best-fit power laws to the observed [OIII] surface brightness profile. Both \citet{fraquelli2003} and \citet{bennert2006} observe NGC 1386 and NGC 5643, and we use the \citet{bennert2006} profiles as these authors remove star formation contamination to [OIII]. For the \citet{fraquelli2003} points, we use their given uncertainties, while the error bars on the \citet{bennert2006} points are propagated from the uncertainties on the power-law indices. We also plot a sample of Type I quasars from \citet[][$z \sim 0.1 - 0.3$]{husemann2013}. These authors provided three measurements of the NLR size: $R_{95}$, the circular radius that encompasses 95\% of the [OIII] flux; $R_{\mathrm{e}}$, the weighted mean of projected pixel distances to the center; and $R_{\mathrm{iso}}$, the isophotal radius out to a limiting surface brightness of $2 \times 10^{-16}$ erg s$^{-1}$ cm$^{-2}$ arcsec$^{-2}$. We plot the $R_{\mathrm{iso}}$ values on our figures, although it is important to note that the limiting surface brightness is deeper than what is used for $R_{\mathrm{int}}$ by a factor of $\sim2-3$ when cosmological dimming is taken into account. If we measure the $R_{\mathrm{iso}}$ values for the quasars observed with SALT using the deeper surface brightness limit from \citet{husemann2013}, and do not account for cosmological dimming, we measure sizes that are only 27\% larger. If we correct the \citeauthor{husemann2013} sizes accordingly (lower them by 27\%), the fitting results discussed below do not change significantly within the presented uncertainties. Finally, we plot the best fit relations to the NLR size - luminosity relationship as given by \citet[][slope = $0.22 \pm 0.04$]{greene2011} and \citet[][slope = $0.25 \pm 0.02$]{liu2013} with grey and black lines. The addition of the SALT sample does not significantly affect these slope measurements. For all samples, including our own, we do not plot the merging AGN sources, where a measurement of the size is far less certain. 
	
	The objects we observed with SALT are consistent with the previously-measured NLR size - $L_{[\mathrm{OIII}]}$ relationship, and while the measurements of the sizes are lower at a given $L_{[\mathrm{OIII}]}$ than what is seen with the \citet{greene2011} and \citet{liu2013} relationships, they agree within 1$\sigma$ of the estimate of the size error. These measurements support the claim that for obscured quasars and nearby Seyfert 2 galaxies, the relationship between NLR size and $L_{[\mathrm{OIII}]}$ is fairly shallow. We also find an agreement in the relationship between the Type I objects from \citet{husemann2013} and the Type II quasars, although there are fewer observed detections of extended emission line regions observed in Type I quasars, as discussed in \citet{liu2013}. 

\begin{deluxetable}{lc}
\tabletypesize{\scriptsize}
\tablecaption{NLR Size \label{tab:NLRsize}}
\tablewidth{0pt}
\tablehead{
\colhead{SDSS Name} & \colhead{log($R_{\mathrm{int}}$ / pc)}  %\\
}
\startdata

J024940.2--082804.6 & $3.7\pm0.2$\phm{$^{a}$} \\
J031428.3--072517.8 & $3.6\pm0.2$\phm{$^{a}$} \\
J033606.7--000754.8 & $3.8\pm0.2$\phm{$^{a}$} \\
J081125.8+073235.4 (P.A. = 130$^{\circ}$) & $3.8\pm0.2$\phm{$^{a}$} \\
J081125.8+073235.4 (P.A. = 220$^{\circ}$) & $3.9\pm0.2$\phm{$^{a}$} \\
J084107.1+033441.3 (P.A. = 240$^{\circ}$) & $3.7\pm0.2$\phm{$^{a}$} \\
J084107.1+033441.3 (P.A. = 140$^{\circ}$) & $3.8\pm0.2$\phm{$^{a}$} \\
J084135.1+10156.3 & -$^{a}$ \\
J110012.4+084616.4 & $3.8\pm0.2$\phm{$^{a}$} \\
J122217.9--000743.8 & -$^{a}$ 

\enddata

\tablenotetext{a}{This object is a double continuum source, and we are unable to use the method outlined in \S \ref{sec:measurements} to flux calibrate the observed spatial profile and estimate R$_{\mathrm{int}}$.}

\end{deluxetable}

\subsection{The Relationship between NLR size and $L_{8 \mu \mathrm{m}}$ }
\label{sec:nlrsizelir}

	\begin{figure}[htbp]
	\epsscale{1.2} 
	\plotone{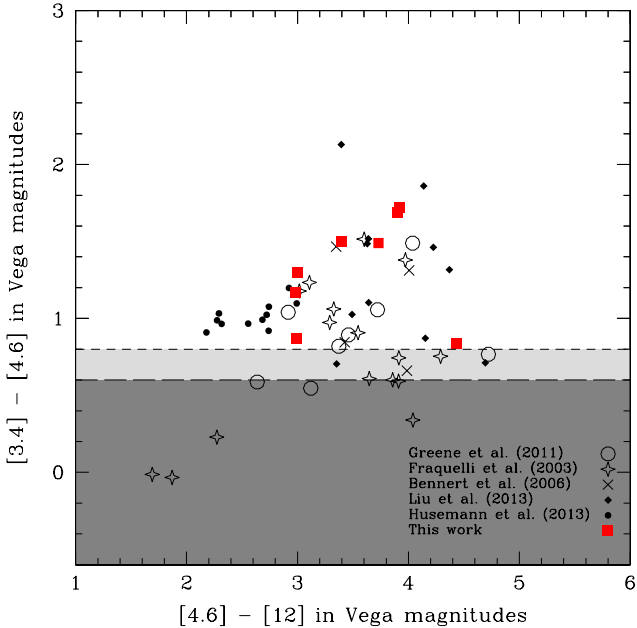} 
	\caption{AGN color-color plot using WISE \citep{wright2010}. AGNs are identified by having red $[3.4]-[4.6]$ colors, and we show the \citet{stern2012} AGN selection division of $[3.4]-[4.6] = 0.8$ with a short dashed line, and shade the region below with light grey. We also show a more relaxed division of $[3.4]-[4.6] = 0.6$ with a long dashed line, and shade the region below with dark grey. Bluer WISE $[3.4]-[4.6]$ colors may indicate contamination in the mid-IR from stellar processes, and we will note objects in the grey or dark grey regions in Figure \ref{fig:NLRsizeIR}. Based on the errors from the WISE photometry, the average uncertainty on the colors for the points is 0.05.
	\label{fig:WISEcolors}} 
	\epsscale{1.}
         \end{figure}
	
	$L_{[\mathrm{OIII}]}$ is often used as a proxy for intrinsic AGN power \citep[e.g.][]{heckman2005}, but [OIII] emission can be strongly affected by dust extinction and NLR properties \citep{baskin2005,wild2011}. In order to examine the intrinsic AGN luminosity that is not dependent on extinction or NLR characteristics, we can use infrared luminosities for those objects that are dominated by emission from the AGN. Using WISE All-Sky Survey data, we have collected the mid-IR photometry for our sample and all of the objects from the literature plotted in Figure \ref{fig:NLRsize}. We first examine the full sample of objects to see which may have contamination in the mid-IR from stellar processes. We plot the WISE $[3.4] - [4.6]$ vs $[4.6] - [12]$ colors\footnote{All WISE magnitudes are in Vega magnitudes.} for the full sample in Figure \ref{fig:WISEcolors}. In this plot, we show the \citet{stern2012} demarcation $[3.4] - [4.6] \geq 0.8$ that these authors used to separate AGNs from non-AGNs. Emission from hot dust near the central black hole in an AGN produces characteristic red mid-IR colors that can be used to identify AGNs. The objects in this paper, along with the bulk of the objects from \citet{bennert2006}, \citet{greene2011}, \citet{liu2013}, and \citet{husemann2013} have colors that indicate AGN activity as described in both \citet{wright2010} and \citet{stern2012}. Many of the local Seyfert 2 galaxies from \citet{fraquelli2003}, however, lie below the \citeauthor{stern2012} demarcation, and we use this diagram to indicate that while the bulk of the objects we analyze are dominated by the AGN component, those with bluer IR colors may be contaminated in their WISE photometry by IR emission from stellar processes. 

	\begin{figure*}[htbp]
	\epsscale{0.9} 
	\plotone{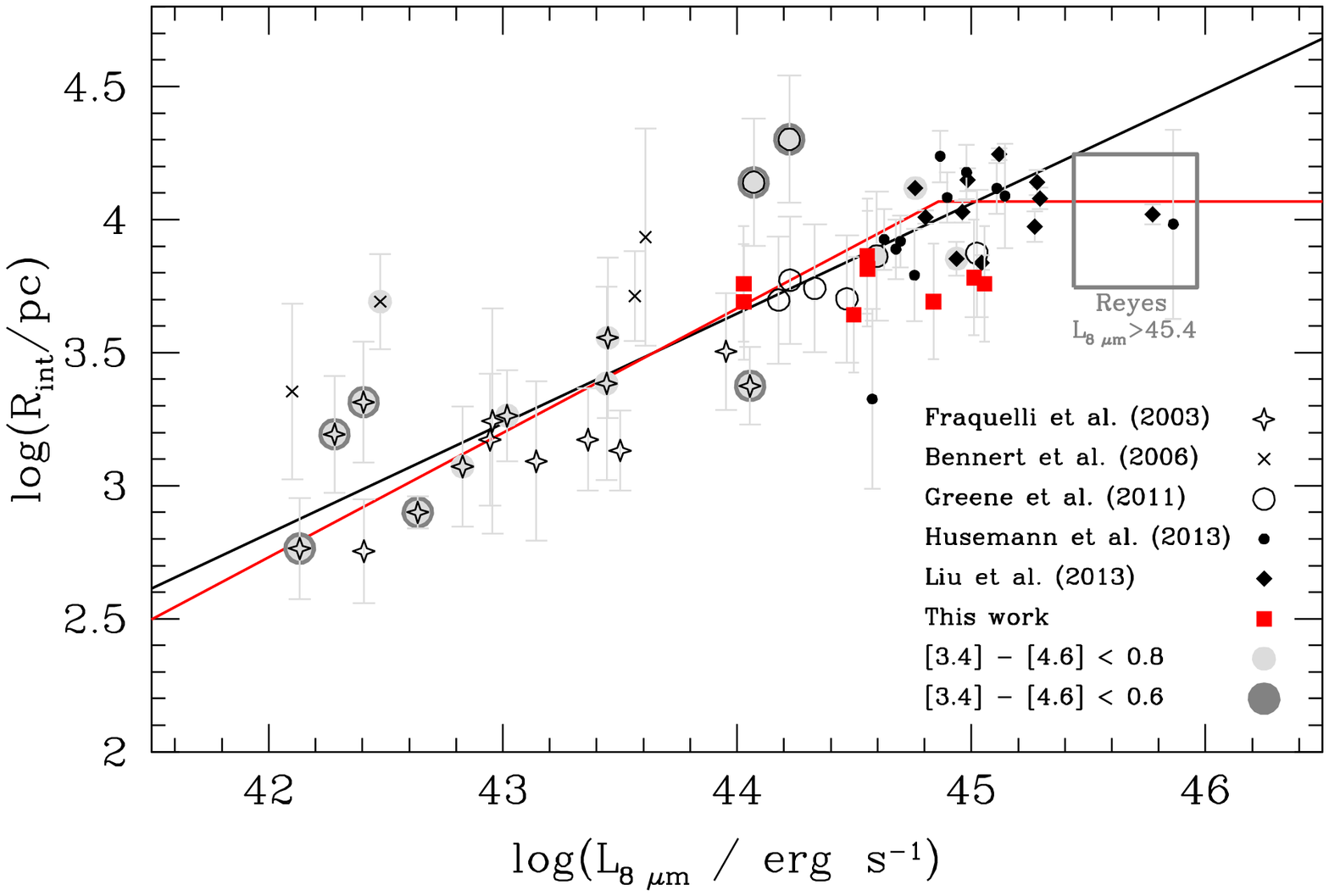} 
	\caption{Radii of the [OIII]$\lambda$5007-emitting region plotted against the interpolated $L_{8\mu \mathrm{m}}$ estimated from WISE photometry. The symbols are the same as in Figure \ref{fig:NLRsize}. We plot those objects from Figure \ref{fig:WISEcolors} with WISE color $[3.4]-[4.6] < 0.8$ superimposed on a light grey circle, and those objects with WISE color $[3.4]-[4.6] < 0.6$ are plotted over a dark grey circle, to indicate which objects may be suffering from contamination by stellar processes. We see a relationship between $L_{8\mu\mathrm{m}}$ and the size of the NLR, and we see evidence for a flattening of the relationship at high $L_{8\mu\mathrm{m}}$. We plot the linear fit for every point except the two highest luminosity quasars with a black line (slope = $0.41 \pm 0.02$). We plot a piecewise linear fit with all of the points including the two highest luminosity quasars with a red line (slope = $0.47 \pm 0.02$) to demonstrate the existence of the flattening of the relationship at $\log{(R_{\mathrm{max}}/\mathrm{pc})} = 4.07$. See the text in \S \ref{sec:nlrsizelir} for a description of these fits. We also show the area occupied by the IR-bright quasars from the sample of \citet{reyes2008} with a dark grey box, where the sizes have been inferred from their values of $L_{[\mathrm{OIII}]}$ and the relationship in Figure \ref{fig:NLRsize}. 
	\label{fig:NLRsizeIR}} 
	\epsscale{1.}
         \end{figure*}
	
	We estimate the rest-frame IR luminosity using the WISE [4.6], [12], and [22] bands, which were interpolated in log-log space to find the flux and luminosity at rest-frame 8 $\mu$m ($L_{8\mu \mathrm{m}}$) for each object. We model the AGN mid-IR emission with a simple power-law, and for this analysis, we do not account for the individual filter response functions and assume that the flux is measured at the central wavelength for each filter. Based on the WISE colors for these objects, we estimate that any flux corrections would be on the order of a few percent \citep{wright2010}, so they are not included here. We use the luminosity at rest-frame $8\mu$m, corresponding to an observed-frame wavelength of $\sim 12 \mu$m, which should be dominated by the warm and hot dust heated by the AGN. We plot the NLR size - IR luminosity relationship in Figure \ref{fig:NLRsizeIR}. Those objects with IR emission potentially contaminated by star formation ($[3.4] - [4.6] < 0.8$) as indicated by Figure \ref{fig:WISEcolors} are plotted over light grey circles. We also mark those objects with $[3.4] - [4.6] < 0.6$ with dark grey circles. As in Figure \ref{fig:NLRsize}, we see a strong relationship such that AGNs with larger 8 $\mu$m luminosities have larger NLR sizes. However, at the high $L_{8\mu \mathrm{m}}$ end of the relation, two objects, SDSS J085829.59+441734.7 from the \citet{liu2013} sample, and PG 1700+518 from the \citet{husemann2013} sample, both have $\sim 10$ kpc NLRs, which hints at the existence of a flattening of the relation at high $L_{8\mu \mathrm{m}}$.

As there are only two objects, these high $L_{8\mu \mathrm{m}}$ quasars may be statistical outliers. We removed these objects from our sample and fit the data with a linear least squares regression of the form:

\begin{equation}
 \log{(R_{\mathrm{int}} / \mathrm{pc})} = \alpha \times \log{L^{44}_{8 \mu \mathrm{m}} } + \beta
\end{equation}

where $L^{44}_{8 \mu \mathrm{m}} =  (L_{8 \mu \mathrm{m}}/10^{44}\, \mathrm{erg}\, \mathrm{s}^{-1})$. We report the best-fitting parameters, as well as the reduced $\chi^2$ for the fit, without these high $L_{8 \mu \mathrm{m}}$ objects in the bottom three rows of Table \ref{tab:bestfit}. We also provide the parameters for fits without blue WISE $[3.4] - [4.6]$ colors. We plot the linear fit without the two high $L_{8 \mu \mathrm{m}}$ objects with a black line in Figure \ref{fig:NLRsizeIR}. The slope of the relation stays largely the same, with $\alpha \approx 0.4$, both with and without the inclusion of objects with bluer WISE $[3.4] - [4.6]$ colors. We also use a linear fit to the data including the two high $L_{8 \mu \mathrm{m}}$ objects and report the best-fitting parameters in the middle three rows of Table \ref{tab:bestfit}. The best-fit slope for a linear fit including the two IR-bright quasars is fairly shallow compared to the fit without the points, but the goodness-of-fit is significantly worse as indicated by the reduced $\chi^2$ values. 

The position of the two high-luminosity points, however, may be indicative of a flattening of the relation at the high luminosity end. In this scenario, we would be seeing evidence that the size of the NLR in the most luminous quasars has an upper limit. To model this flattening, we fit the data using a piecewise linear relationship:

\begin{equation}\label{eq:piece}
 \log{(R_{\mathrm{int}} / \mathrm{pc})} = \left\{ 
  \begin{array}{l l}
     \alpha \times \log{L^{44}_{8 \mu \mathrm{m}} } + \beta & \quad L_{8 \mu \mathrm{m}} < {L}_{0} \\
      \log{(R_{\mathrm{0}}/\mathrm{pc})} & \quad L_{8 \mu \mathrm{m}} \geq {L}_{0} \\  

  \end{array} \right.
\end{equation}

In this fit, ${L}_{0}$ is the turnover luminosity, and $\log{(R_{\mathrm{0}}/\mathrm{pc})} = \alpha \times \log{{L}_{0}} + \beta$, the NLR size at the turnover luminosity. We provide the best-fitting values for this piecewise relation in the first three rows from Table \ref{tab:bestfit}. We plot the piecewise fit to all of the data with a red line in Figure \ref{fig:NLRsizeIR}. The slope of the rising portion of the relationship is similar to but marginally steeper than what we observe when we fit without the two highest luminosity objects. The inclusion of objects with bluer $[3.4]-[4.6]$ colors does not significantly affect the fit parameters. We also fit the data using Equation \ref{eq:piece}, but we allow the slope of the second linear function to be a free parameter, and found that this best-fit slope was consistent with zero. Together, we find that the slope for the rising portion of the relation is $\alpha = 0.4 - 0.5$, with a flattening at a characteristic luminosity of $\log{(L_{0} / \mathrm{erg}\, \mathrm{s}^{-1})} = 44.8 - 45.1$, and an NLR size of around $\log{(R_{\mathrm{0}}/\mathrm{pc})} = 4.07 - 4.10$ ($\sim 12$ kpc). It is important to note that PG 1700+518, one of the two high $L_{8 \mu \mathrm{m}}$ objects, is a broad absorption line quasar with a nearby companion \citep{stockton1998, evans2009}. It has been proposed that PG 1700+518 has undergone a collision with the companion galaxy \citep{hines1999}, which makes a measurement of the true size of the NLR difficult. This is reflected in the large uncertainty on the NLR size measurement. At the same time, based on the WISE $[3.4] - [4.6]$ colors for this object, the origin of the large $L_{8\mu \mathrm{m}}$ is most likely dominated by AGN activity, regardless of triggering mechanism. We still choose to include it in our sample, although, due to this large uncertainty, if we remove this object we obtain the same best-fit parameters on our piecewise fitting. We will further discuss the implications of a potential flattening in this relationship in the next section.
	
\begin{deluxetable*}{lccccc}
\tabletypesize{\scriptsize}
\tablecaption{$L_{8\mu \mathrm{m}}$ vs. $R_{\mathrm{int}}$ Best Fit Properties \label{tab:bestfit}}
\tablewidth{0pt}
\tablehead{
\colhead{Sample} & \colhead{$\alpha$} & \colhead{$\beta$} & \colhead{$\log$($L_{0}$/erg s$^{-1}$)} & \colhead{$\log$($R_{0}$/pc)} & \colhead{$\chi^2_{\mathrm{red}}$}
}
\startdata

All Objects &  $0.47 \pm 0.02$ & $3.67 \pm 0.02$ & $44.86 \pm 0.05$ & $4.07 \pm 0.03$ & 4.48 \\
$[3.4] - [4.6] > 0.8$ &  $0.44 \pm 0.03$ & $3.61 \pm 0.03$ & $45.12 \pm 0.06$ & $4.10 \pm 0.06$ & 4.22 \\
$[3.4] - [4.6] > 0.6$ &  $0.42 \pm 0.03$ & $3.70 \pm 0.03$ & $44.87 \pm 0.05$ & $4.07 \pm 0.05$ & 4.62 \\

All Objects &  $0.36\pm0.02$ & $3.68 \pm 0.02$ & - & - & 5.79 \\
$[3.4] - [4.6] > 0.8$ &  $0.28 \pm 0.03$ & $3.75 \pm 0.03$ & - & - & 5.49 \\
$[3.4] - [4.6] > 0.6$ &  $0.26 \pm 0.03$ & $3.78 \pm 0.02$ & - & - & 5.42 \\
\hline

\hline
\hline

All Objects\tablenotemark{a} &  $0.41\pm0.02$ & $3.65 \pm 0.02$ & - & - & 4.61 \\
$[3.4] - [4.6] > 0.8\tablenotemark{a}$ &  $0.41 \pm 0.03$ & $3.64 \pm 0.03$ & - & - & 4.38 \\
$[3.4] - [4.6] > 0.6\tablenotemark{a}$ &  $0.35 \pm 0.03$ & $3.71 \pm 0.03$ & - & - & 4.60 

\enddata

\tablenotetext{a}{This linear fit was done without the two high $L_{8\mu \mathrm{m}}$ sources.}

\end{deluxetable*}

\section{Discussion and Conclusions}
\label{sec:discussion}

We have used luminous, obscured quasars to examine the effect that bright active nuclei are having on the gas in their host galaxies at large radii. Our SALT RSS long-slit results strengthen the conclusions made in previous studies, and primarily indicate that the accreting black holes in the galactic centers of these objects are  ionizing gas out to large radii. 

We examine the relationship between the NLR size and $L_{8\mu \mathrm{m}}$, which serves as a more direct indicator of the current nuclear power independent of NLR properties. We see a strong power-law trend such that more luminous AGNs are serving to ionize larger narrow-line regions. This relation goes as approximately $R_{\mathrm{int}} \sim L_{8\mu\mathrm{m}}^{1/2}$, steeper than the observed relationship between $R_{\mathrm{int}}$ and $L_{[\mathrm{OIII}]}$. We also observe hints of a flattening in the $R_{\mathrm{int}}$ versus $L_{8\mu\mathrm{m}}$ relationship at around $\sim12$ kpc for the most luminous AGNs. This suggestion of a flattening of this relationship could be seen as an upper limit on the amount of available gas to be ionized. At high luminosities, the NLR may be completely ionized at large radii, and any increase in the AGN luminosity, as traced by $L_{8\mu\mathrm{m}}$, would only then serve to produce more ionizing photons which escape the galaxy. 

A flattening of the relationship between AGN luminosity and NLR size for the most powerful quasars has been suggested to exist by multiple authors, including \citet{netzer2004} and \citet{greene2011}. \citet{netzer2004} examined a sample of high-redshift AGNs and argued that the relationship between the NLR size and $L_{[\mathrm{OIII}]}$ must break down for very high luminosity sources, where the relationship predicts unphysically large sizes. Using the observed slope seen in Figure \ref{fig:NLRsize}, a source with log($L_{[\mathrm{OIII}]}/\mathrm{erg\, s}^{-1}) = 45$ would have a NLR size of $\sim 30$ kpc.  The objects that comprise our sample, as well as the samples from \citet{greene2011} and \citet{liu2013}, are chosen to have the largest values of $L_{[\mathrm{OIII}]}$ from the \citet{zakamska2003} and \citet{reyes2008} obscured quasar samples. As there are very few quasars in the local universe with larger [OIII] luminosities, it is likely that we are indeed seeing the maximum physical extent of NLRs in these types of objects. 

\begin{figure}[htbp]
	\epsscale{1.2} 
	\plotone{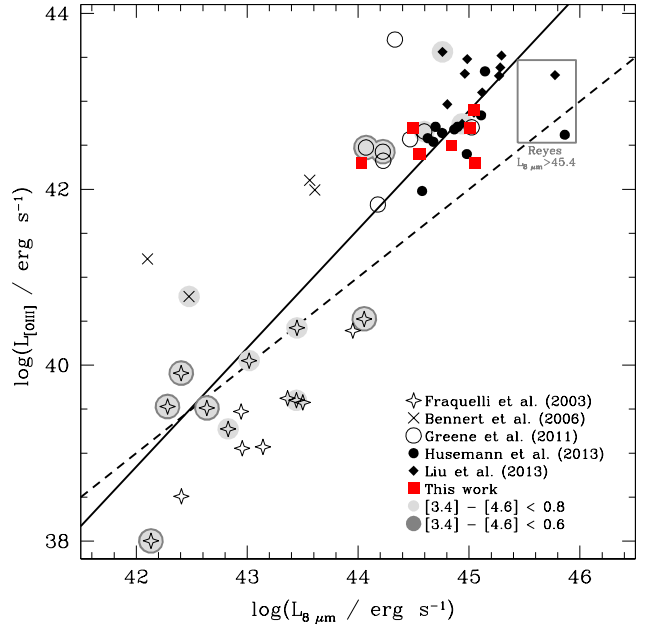} 
	\caption{$L_{[\mathrm{OIII}]}$ plotted against $L_{8\mu \mathrm{m}}$ estimated from WISE photometry. The symbols are the same as in Figures \ref{fig:NLRsize} and \ref{fig:NLRsizeIR}. We plot a linear relationship between the two variables with a dashed line, and a fit to the entire dataset (slope = 1.35) with a solid line. These results provide evidence that $L_{[\mathrm{OIII}]}$ is not linearly correlated with the nuclear luminosity, which may be better probed by $L_{8\mu\mathrm{m}}$. We also show the area occupied by the IR-bright quasars from the sample of \citet{reyes2008} with a dark grey box. 
	\label{fig:IROIII}} 
	\epsscale{1.}
         \end{figure}

To further understand the observed relations between $R_{\mathrm{int}}$ and $L_{[\mathrm{OIII}]}$ and $L_{8\mu\mathrm{m}}$, we consider the simplest possible model for the NLR: a finite sphere of gas with constant density $n$ surrounding the accreting black hole, or, more realistically, constant density clouds contained within a lower density medium with a constant filling factor. In these scenarios, the ionization parameter $U$ at a radius $r$ from the nucleus depends on the central ionizing luminosity $L$ as well as the density of the gas clouds: $U \propto L / r^2 n^2$. The size of the ionized region will be limited by the radius at which the ionization parameter is above some threshold for producing a sufficient density of [OIII]. In this case, the radius of the NLR scales as $L^{1/2}$. This simple prediction is very close to our measured slope of $0.47 \pm 0.02$ in the relationship between log($L_{8 \mu \mathrm{m}}$) and log($R_{\mathrm{int}}$) (Figure \ref{fig:NLRsizeIR}). We contrast this with the shallow slope of $0.25$ for the relationship between log($L_{[\mathrm{OIII}]}$) and log($R_{\mathrm{int}}$) seen in Figure \ref{fig:NLRsize} \citep{liu2013}. In our simple picture, the luminosity due to recombination (for example, $L_{\mathrm{H}\beta}$) is dependent on the volume of ionized gas. $L_{\mathrm{H}\beta}$ can be used as a proxy for $L_{[\mathrm{OIII}]}$ as there is evidence in Type II quasars that the ratio of $[\mathrm{OIII}] /\mathrm{H}\beta$ is constant out to several kpc \citep[][Hainline et al. in prep]{liu2013}. For the \citeauthor{liu2013} sample, the $[\mathrm{OIII}] /\mathrm{H}\beta$ ratio does not vary out to radii larger than the typical PSF, and the assumption of a constant line ratio is unaffected by seeing effects. If the number density of the clouds $\zeta$ is constant, and the ionization fraction is near unity, then $L_{\mathrm{H}\beta} \propto L_{[\mathrm{OIII}]} \propto \zeta r^3 n^2$ and the NLR radius then scales as $L^{1/3}_{[\mathrm{OIII}]}$, close to (although slightly steeper than) the slope we observe. This picture would suggest that for powerful quasars, $L_{[\mathrm{OIII}]}$ is not linearly correlated with the nuclear luminosity, which is better probed by more direct measurements such as $L_{8\mu\mathrm{m}}$. We plot the relationship between $L_{[\mathrm{OIII}]}$ and $L_{8\mu\mathrm{m}}$ in Figure \ref{fig:IROIII}. We show a linear relationship between the two quantities with a dashed line. When we fit all of the data together using a linear least squares regression, we obtain a slope of $1.35$ (shown by the black line). Combining our simple models from above, we obtain $L^{1/3}_{[\mathrm{OIII}]} \propto R_{\mathrm{NLR}} \propto L^{1/2}_{\mathrm{8\mu\mathrm{m}}}$ and therefore we predict that $L_{[\mathrm{OIII}]} \propto L^{3/2}_{\mathrm{8\mu\mathrm{m}}}$. The measured slope is shallower, but very close to, the slope predicted by our simple model. We note that by removing the points with bluer WISE $[3.4] - [4.6]$ colors, as well as one highly discrepant point from Bennert et al. (2006) at low $L_{8\mu\mathrm{m}}$, the best-fit slope becomes $1.49$, which agrees very closely with our prediction. We also see a truncation of the assembled data at the high $L_{[\mathrm{OIII}]}$ end, which is naturally explained if we are indeed seeing a maximum size for the NLR.

This simple model also naturally explains the observed flattening of the relationship at large luminosity, since the NLR size is limited by the finite extent of the gas clouds. The turnover size for the NLRs in Figure \ref{fig:NLRsizeIR} agrees well with the quasar host galaxy sizes of $\sim10$ kpc estimated from deep imaging campaigns \citep{mclure1999, schade2000, dunlop2003, kotilainen2013} and is consistent with continuum sizes measured for long-slit spectroscopy of obscured quasars \citep{greene2011}.

This picture is of course highly simplified and does not account for complications such as variations in density or the [OIII]/H$\beta$ ratio. As discussed above, \citet{liu2013} observe a constant [OIII]/H$\beta$ ratio to large radii for Type 2 quasars, with a falloff further out, and propose a simple constant-density model (``Model 1'') similar to the model we describe above, for which the falloff in [OIII]/H$\beta$ is caused by the decline of the ionization parameter with radius. However, this model requires constant pressure in the NLR clouds at all radii, in conflict with theoretical studies and observations of galactic winds. Therefore, \citet{liu2013} prefer a second model where the density and pressure in NLR clouds drops as $r^{-2}$ and the ionization parameter is thus constant with radius. In this case the observed size of the NLR may be limited by the radius at which the gas density is low enough that the clouds are ionized to much higher levels than required to produce a large amount of [OIII] emission. It is not immediately clear how $R_{\mathrm{int}}$ would vary with nuclear luminosity and $L_{[\mathrm{OIII}]}$ in this scenario, and we reserve this more complicated modeling for future work. 

Our results imply that for quasars with $L_{8 \mu \mathrm{m}}$ larger than the turnover luminosity, neither the NLR size nor $L_{[\mathrm{OIII}]}$ will increase further. However, this suggestion would be strengthened by a detailed study of more IR-luminous obscured quasars. We can examine the full \citet{reyes2008} Type II quasar sample to see how representative the IR-bright objects in our assembled sample are. We were able to match 876 out of the 887 quasars in the \citet{reyes2008} sample against the WISE catalogue, and for these matched quasars we estimated $L_{8 \mu \mathrm{m}}$ using the method described above. We find 150 objects ($\sim 17 \%$ of the total WISE-detected sample) with $\log(L_{8\mu \mathrm{m}}) > 44.85$, which represents the turnover luminosity in the relationship seen in Figure \ref{fig:NLRsizeIR}. We calculate $\langle \log{L_{([\mathrm{OIII}]}/\mathrm{erg\, s}^{-1})} \rangle = 42.7 \pm 0.5$ for these IR-bright quasars, which we compare to $\langle \log{(L_{[\mathrm{OIII}]}/\mathrm{erg\, s}^{-1})} \rangle = 43.1 \pm 0.3$ for the quasars in our assembled sample. If we further restrict the \citeauthor{reyes2008} sample to only those with $\log(L_{8\mu \mathrm{m}}) > 45.4$, we find 29 quasars, with $\langle L_{[\mathrm{OIII}]}/\mathrm{erg\, s}^{-1} \rangle = 43.0 \pm 0.5$. These results support our hypothesis that there exists a large population of AGNs with large $L_{8\mu \mathrm{m}}$ but at similar $L_{[\mathrm{OIII}]}$ to those of the two IR-bright sources in our assembled sample. We can also make a prediction for the NLR size of these objects from their $L_{[\mathrm{OIII}]}$ using the relation from \citet{liu2013}: $R_{\mathrm{int}} = 9 \pm 2$ kpc for the quasars with $\log(L_{8\mu \mathrm{m}}) > 44.85$, and $R_{\mathrm{int}} = 10 \pm 3$ kpc for the quasars with $\log(L_{8\mu \mathrm{m}}) > 45.4$.  In Figures \ref{fig:NLRsizeIR} and \ref{fig:IROIII}, we show the area occupied by the IR-bright quasars from the \citeauthor{reyes2008} sample with a dark grey box.  Observations of these objects with future spectroscopic campaigns can definitively test the presence of an upper limit on $R_{\mathrm{int}}$ for these luminous quasars. We note that it would be fundamentally interesting if these objects have larger sizes than what is predicted by the flattening in the $R_{\mathrm{int}}$ vs. $L_{8\mu \mathrm{m}}$ relationship, as this would imply a different relationship between NLR size and $L_{\mathrm{[OIII]}}$. This would indicate that the highest luminosity quasars have significantly different gas reservoirs than their lower-luminosity counterparts, which could have implications for models of quasar fueling.

To conclude, we have used SALT RSS long-slit spectroscopy to examine the sizes of the NLRs of a sample of 8 obscured quasars. Our primary results are the following:

\begin{enumerate}
  \item The NLR sizes from the objects in our quasar sample are similar to what has been measured in other studies of these objects, and help define the relationship between NLR size and $L_{[\mathrm{OIII}]}$ first discussed in \citet{bennert2002}. Our objects are consistent with a shallow slope in this relation as described in \citet{greene2011} and \citet{liu2013}.
  \item Using WISE mid-IR photometry, we have estimated the rest-frame $L_{8\mu \mathrm{m}}$ for our quasar sample, as well as samples from the literature with WISE coverage. We find that NLR size goes as approximately $L^{1/2}_{8\mu\mathrm{m}}$, as expected from the simple scenario of a constant-density cloud illuminated by a central ionizing source. We also observe a possible flattening of the relationship between NLR size and AGN luminosity for quasars with $\log{(L_{8\mu\mathrm{m}} / \mathrm{erg}\, \mathrm{s}^{-1})} > 44.8 - 44.9$, where their sizes plateau at $\sim 12$ kpc. While we only observe this flattening of the relationship with two objects, the trend could indicate that we are seeing the maximum size of the NLR in these galaxies, beyond which the ionization state and density of the gas is such that excess ionizing photons escape the galaxy. We furthermore highlight the many additional IR-luminous targets in the full sample of Type II quasars from \citet{reyes2008}, which would be ideal for a follow-up exploration of this turnover. 
   
\end{enumerate}
  
\acknowledgments 

The authors wish to thank Alice Shapley for a helpful discussion regarding the SALT data reduction. KNH, RCH and ADM were partially supported by NASA through ADAP award NNX12AE38G and by the National Science Foundation through grant numbers 1211096 and 1211112.

\bibliographystyle{apj}

\begin{thebibliography}{41}
\expandafter\ifx\csname natexlab\endcsname\relax\def\natexlab#1{#1}\fi

\bibitem[{{Asmus} {et~al.}(2011){Asmus}, {Gandhi}, {Smette}, {H{\"o}nig}, \&
  {Duschl}}]{asmus2011}
{Asmus}, D., {Gandhi}, P., {Smette}, A., {H{\"o}nig}, S.~F., \& {Duschl}, W.~J.
  2011, \aap, 536, A36

\bibitem[{{Baskin} \& {Laor}(2005)}]{baskin2005}
{Baskin}, A. \& {Laor}, A. 2005, \mnras, 358, 1043

\bibitem[{{Becker} {et~al.}(1995){Becker}, {White}, \& {Helfand}}]{becker1995}
{Becker}, R.~H., {White}, R.~L., \& {Helfand}, D.~J. 1995, \apj, 450, 559

\bibitem[{{Bennert} {et~al.}(2002){Bennert}, {Falcke}, {Schulz}, {Wilson}, \&
  {Wills}}]{bennert2002}
{Bennert}, N., {Falcke}, H., {Schulz}, H., {Wilson}, A.~S., \& {Wills}, B.~J.
  2002, \apjl, 574, L105

\bibitem[{{Bennert} {et~al.}(2006){Bennert}, {Jungwiert}, {Komossa}, {Haas}, \&
  {Chini}}]{bennert2006}
{Bennert}, N., {Jungwiert}, B., {Komossa}, S., {Haas}, M., \& {Chini}, R. 2006,
  \aap, 456, 953

\bibitem[{{Dunlop} {et~al.}(2003){Dunlop}, {McLure}, {Kukula}, {Baum}, {O'Dea},
  \& {Hughes}}]{dunlop2003}
{Dunlop}, J.~S., {McLure}, R.~J., {Kukula}, M.~J., {Baum}, S.~A., {O'Dea},
  C.~P., \& {Hughes}, D.~H. 2003, \mnras, 340, 1095

\bibitem[{{Evans} {et~al.}(2009){Evans}, {Hines}, {Barthel}, {Schneider},
  {Surace}, {Sanders}, {Vavilkin}, {Frayer}, {Tacconi}, \&
  {Storrie-Lombardi}}]{evans2009}
{Evans}, A.~S., {Hines}, D.~C., {Barthel}, P., {Schneider}, G., {Surace},
  J.~A., {Sanders}, D.~B., {Vavilkin}, T., {Frayer}, D.~T., {Tacconi}, L.~J.,
  \& {Storrie-Lombardi}, L.~J. 2009, \aj, 138, 262

\bibitem[{{Fraquelli} {et~al.}(2003){Fraquelli}, {Storchi-Bergmann}, \&
  {Levenson}}]{fraquelli2003}
{Fraquelli}, H.~A., {Storchi-Bergmann}, T., \& {Levenson}, N.~A. 2003, \mnras,
  341, 449

\bibitem[{{Greene} {et~al.}(2011){Greene}, {Zakamska}, {Ho}, \&
  {Barth}}]{greene2011}
{Greene}, J.~E., {Zakamska}, N.~L., {Ho}, L.~C., \& {Barth}, A.~J. 2011, \apj,
  732, 9

\bibitem[{{Heckman} {et~al.}(2005){Heckman}, {Ptak}, {Hornschemeier}, \&
  {Kauffmann}}]{heckman2005}
{Heckman}, T.~M., {Ptak}, A., {Hornschemeier}, A., \& {Kauffmann}, G. 2005,
  \apj, 634, 161

\bibitem[{{Hines} {et~al.}(1999){Hines}, {Low}, {Thompson}, {Weymann}, \&
  {Storrie-Lombardi}}]{hines1999}
{Hines}, D.~C., {Low}, F.~J., {Thompson}, R.~I., {Weymann}, R.~J., \&
  {Storrie-Lombardi}, L.~J. 1999, \apj, 512, 140

\bibitem[{{Horst} {et~al.}(2008){Horst}, {Gandhi}, {Smette}, \&
  {Duschl}}]{horst2008}
{Horst}, H., {Gandhi}, P., {Smette}, A., \& {Duschl}, W.~J. 2008, \aap, 479,
  389

\bibitem[{{Husemann} {et~al.}(2013){Husemann}, {Wisotzki}, {S{\'a}nchez}, \&
  {Jahnke}}]{husemann2013}
{Husemann}, B., {Wisotzki}, L., {S{\'a}nchez}, S.~F., \& {Jahnke}, K. 2013,
  \aap, 549, A43

\bibitem[{{Kobulnicky} {et~al.}(2003){Kobulnicky}, {Willmer}, {Phillips},
  {Koo}, {Faber}, {Weiner}, {Sarajedini}, {Simard}, \& {Vogt}}]{kobulnicky2003}
{Kobulnicky}, H.~A., {Willmer}, C.~N.~A., {Phillips}, A.~C., {Koo}, D.~C.,
  {Faber}, S.~M., {Weiner}, B.~J., {Sarajedini}, V.~L., {Simard}, L., \&
  {Vogt}, N.~P. 2003, \apj, 599, 1006

\bibitem[{{Kotilainen} {et~al.}(2013){Kotilainen}, {Falomo}, {Bettoni},
  {Karhunen}, \& {Uslenghi}}]{kotilainen2013}
{Kotilainen}, J., {Falomo}, R., {Bettoni}, D., {Karhunen}, K., \& {Uslenghi},
  M. 2013, ArXiv e-prints

\bibitem[{{Krabbe} {et~al.}(2001){Krabbe}, {B{\"o}ker}, \&
  {Maiolino}}]{krabbe2001}
{Krabbe}, A., {B{\"o}ker}, T., \& {Maiolino}, R. 2001, \apj, 557, 626

\bibitem[{{Kreimeyer} \& {Veilleux}(2013)}]{kreimeyer2013}
{Kreimeyer}, K. \& {Veilleux}, S. 2013, \apjl, 772, L11

\bibitem[{{Liu} {et~al.}(2013){Liu}, {Zakamska}, {Greene}, {Nesvadba}, \&
  {Liu}}]{liu2013}
{Liu}, G., {Zakamska}, N.~L., {Greene}, J.~E., {Nesvadba}, N.~P.~H., \& {Liu},
  X. 2013, \mnras

\bibitem[{{Lutz} {et~al.}(2004){Lutz}, {Maiolino}, {Spoon}, \&
  {Moorwood}}]{lutz2004}
{Lutz}, D., {Maiolino}, R., {Spoon}, H.~W.~W., \& {Moorwood}, A.~F.~M. 2004,
  \aap, 418, 465

\bibitem[{{Matsuta} {et~al.}(2012){Matsuta}, {Gandhi}, {Dotani}, {Nakagawa},
  {Isobe}, {Ueda}, {Ichikawa}, {Terashima}, {Oyabu}, {Yamamura}, \&
  {Stawarz}}]{matsuta2012}
{Matsuta}, K., {Gandhi}, P., {Dotani}, T., {Nakagawa}, T., {Isobe}, N., {Ueda},
  Y., {Ichikawa}, K., {Terashima}, Y., {Oyabu}, S., {Yamamura}, I., \&
  {Stawarz}, {\L}. 2012, \apj, 753, 104

\bibitem[{{McLure} {et~al.}(1999){McLure}, {Kukula}, {Dunlop}, {Baum}, {O'Dea},
  \& {Hughes}}]{mclure1999}
{McLure}, R.~J., {Kukula}, M.~J., {Dunlop}, J.~S., {Baum}, S.~A., {O'Dea},
  C.~P., \& {Hughes}, D.~H. 1999, \mnras, 308, 377

\bibitem[{{Netzer} {et~al.}(2004){Netzer}, {Shemmer}, {Maiolino}, {Oliva},
  {Croom}, {Corbett}, \& {di Fabrizio}}]{netzer2004}
{Netzer}, H., {Shemmer}, O., {Maiolino}, R., {Oliva}, E., {Croom}, S.,
  {Corbett}, E., \& {di Fabrizio}, L. 2004, \apj, 614, 558

\bibitem[{{Pier} \& {Krolik}(1993)}]{pier1993}
{Pier}, E.~A. \& {Krolik}, J.~H. 1993, \apj, 418, 673

\bibitem[{{Reyes} {et~al.}(2008){Reyes}, {Zakamska}, {Strauss}, {Green},
  {Krolik}, {Shen}, {Richards}, {Anderson}, \& {Schneider}}]{reyes2008}
{Reyes}, R., {Zakamska}, N.~L., {Strauss}, M.~A., {Green}, J., {Krolik}, J.~H.,
  {Shen}, Y., {Richards}, G.~T., {Anderson}, S.~F., \& {Schneider}, D.~P. 2008,
  \aj, 136, 2373

\bibitem[{{Schade} {et~al.}(2000){Schade}, {Boyle}, \& {Letawsky}}]{schade2000}
{Schade}, D.~J., {Boyle}, B.~J., \& {Letawsky}, M. 2000, \mnras, 315, 498

\bibitem[{{Schmitt} {et~al.}(2003){Schmitt}, {Donley}, {Antonucci},
  {Hutchings}, {Kinney}, \& {Pringle}}]{schmitt2003}
{Schmitt}, H.~R., {Donley}, J.~L., {Antonucci}, R.~R.~J., {Hutchings}, J.~B.,
  {Kinney}, A.~L., \& {Pringle}, J.~E. 2003, \apj, 597, 768

\bibitem[{{Smith} {et~al.}(2006){Smith}, {Nordsieck}, {Burgh}, {Percival},
  {Williams}, {O'Donohue}, {O'Connor}, \& {Schier}}]{smith2006}
{Smith}, M.~P., {Nordsieck}, K.~H., {Burgh}, E.~B., {Percival}, J.~W.,
  {Williams}, T.~B., {O'Donohue}, D., {O'Connor}, J., \& {Schier}, J.~A. 2006,
  in Society of Photo-Optical Instrumentation Engineers (SPIE) Conference
  Series, Vol. 6269, Society of Photo-Optical Instrumentation Engineers (SPIE)
  Conference Series

\bibitem[{{Stern} {et~al.}(2012){Stern}, {Assef}, {Benford}, {Blain}, {Cutri},
  {Dey}, {Eisenhardt}, {Griffith}, {Jarrett}, {Lake}, {Masci}, {Petty},
  {Stanford}, {Tsai}, {Wright}, {Yan}, {Harrison}, \& {Madsen}}]{stern2012}
{Stern}, D., {Assef}, R.~J., {Benford}, D.~J., {Blain}, A., {Cutri}, R., {Dey},
  A., {Eisenhardt}, P., {Griffith}, R.~L., {Jarrett}, T.~H., {Lake}, S.,
  {Masci}, F., {Petty}, S., {Stanford}, S.~A., {Tsai}, C.-W., {Wright}, E.~L.,
  {Yan}, L., {Harrison}, F., \& {Madsen}, K. 2012, \apj, 753, 30

\bibitem[{{Stockton}(1976)}]{stockton1976}
{Stockton}, A. 1976, \apjl, 205, L113

\bibitem[{{Stockton} {et~al.}(1998){Stockton}, {Canalizo}, \&
  {Close}}]{stockton1998}
{Stockton}, A., {Canalizo}, G., \& {Close}, L.~M. 1998, \apjl, 500, L121

\bibitem[{{Stockton} \& {MacKenty}(1987)}]{stocktonmackenty1987}
{Stockton}, A. \& {MacKenty}, J.~W. 1987, \apj, 316, 584

\bibitem[{{Trujillo} {et~al.}(2001){Trujillo}, {Aguerri}, {Cepa}, \&
  {Guti{\'e}rrez}}]{trujillo2001}
{Trujillo}, I., {Aguerri}, J.~A.~L., {Cepa}, J., \& {Guti{\'e}rrez}, C.~M.
  2001, \mnras, 328, 977

\bibitem[{{Villar-Mart{\'{\i}}n} {et~al.}(2010){Villar-Mart{\'{\i}}n},
  {Tadhunter}, {P{\'e}rez}, {Humphrey}, {Mart{\'{\i}}nez-Sansigre}, {Delgado},
  \& {P{\'e}rez-Torres}}]{villarmartin2010}
{Villar-Mart{\'{\i}}n}, M., {Tadhunter}, C., {P{\'e}rez}, E., {Humphrey}, A.,
  {Mart{\'{\i}}nez-Sansigre}, A., {Delgado}, R.~G., \& {P{\'e}rez-Torres}, M.
  2010, \mnras, 407, L6

\bibitem[{{Wampler} {et~al.}(1975){Wampler}, {Burbidge}, {Baldwin}, \&
  {Robinson}}]{wampler1975}
{Wampler}, E.~J., {Burbidge}, E.~M., {Baldwin}, J.~A., \& {Robinson}, L.~B.
  1975, \apjl, 198, L49

\bibitem[{{White} {et~al.}(1997){White}, {Becker}, {Helfand}, \&
  {Gregg}}]{white1997}
{White}, R.~L., {Becker}, R.~H., {Helfand}, D.~J., \& {Gregg}, M.~D. 1997,
  \apj, 475, 479

\bibitem[{{Wild} {et~al.}(2011){Wild}, {Groves}, {Heckman}, {Sonnentrucker},
  {Armus}, {Schiminovich}, {Johnson}, {Martins}, \& {Lamassa}}]{wild2011}
{Wild}, V., {Groves}, B., {Heckman}, T., {Sonnentrucker}, P., {Armus}, L.,
  {Schiminovich}, D., {Johnson}, B., {Martins}, L., \& {Lamassa}, S. 2011,
  \mnras, 410, 1593

\bibitem[{{Wright} {et~al.}(2010){Wright}, {Eisenhardt}, {Mainzer}, {Ressler},
  {Cutri}, {Jarrett}, {Kirkpatrick}, {Padgett}, {McMillan}, {Skrutskie},
  {Stanford}, {Cohen}, {Walker}, {Mather}, {Leisawitz}, {Gautier}, {McLean},
  {Benford}, {Lonsdale}, {Blain}, {Mendez}, {Irace}, {Duval}, {Liu}, {Royer},
  {Heinrichsen}, {Howard}, {Shannon}, {Kendall}, {Walsh}, {Larsen}, {Cardon},
  {Schick}, {Schwalm}, {Abid}, {Fabinsky}, {Naes}, \& {Tsai}}]{wright2010}
{Wright}, E.~L., et al. 2010, \aj, 140, 1868

\bibitem[{{Xu} {et~al.}(1999){Xu}, {Livio}, \& {Baum}}]{xu1999}
{Xu}, C., {Livio}, M., \& {Baum}, S. 1999, \aj, 118, 1169

\bibitem[{{York} {et~al.}(2000){York}, {Adelman}, {Anderson}, {Anderson},
  {Annis}, {Bahcall}, {Bakken}, {Barkhouser}, {Bastian}, {Berman}, {Boroski},
  {Bracker}, {Briegel}, {Briggs}, {Brinkmann}, {Brunner}, {Burles}, {Carey},
  {Carr}, {Castander}, {Chen}, {Colestock}, {Connolly}, {Crocker}, {Csabai},
  {Czarapata}, {Davis}, {Doi}, {Dombeck}, {Eisenstein}, {Ellman}, {Elms},
  {Evans}, {Fan}, {Federwitz}, {Fiscelli}, {Friedman}, {Frieman}, {Fukugita},
  {Gillespie}, {Gunn}, {Gurbani}, {de Haas}, {Haldeman}, {Harris}, {Hayes},
  {Heckman}, {Hennessy}, {Hindsley}, {Holm}, {Holmgren}, {Huang}, {Hull},
  {Husby}, {Ichikawa}, {Ichikawa}, {Ivezi{\'c}}, {Kent}, {Kim}, {Kinney},
  {Klaene}, {Kleinman}, {Kleinman}, {Knapp}, {Korienek}, {Kron}, {Kunszt},
  {Lamb}, {Lee}, {Leger}, {Limmongkol}, {Lindenmeyer}, {Long}, {Loomis},
  {Loveday}, {Lucinio}, {Lupton}, {MacKinnon}, {Mannery}, {Mantsch}, {Margon},
  {McGehee}, {McKay}, {Meiksin}, {Merelli}, {Monet}, {Munn}, {Narayanan},
  {Nash}, {Neilsen}, {Neswold}, {Newberg}, {Nichol}, {Nicinski}, {Nonino},
  {Okada}, {Okamura}, {Ostriker}, {Owen}, {Pauls}, {Peoples}, {Peterson},
  {Petravick}, {Pier}, {Pope}, {Pordes}, {Prosapio}, {Rechenmacher}, {Quinn},
  {Richards}, {Richmond}, {Rivetta}, {Rockosi}, {Ruthmansdorfer}, {Sandford},
  {Schlegel}, {Schneider}, {Sekiguchi}, {Sergey}, {Shimasaku}, {Siegmund},
  {Smee}, {Smith}, {Snedden}, {Stone}, {Stoughton}, {Strauss}, {Stubbs},
  {SubbaRao}, {Szalay}, {Szapudi}, {Szokoly}, {Thakar}, {Tremonti}, {Tucker},
  {Uomoto}, {Vanden Berk}, {Vogeley}, {Waddell}, {Wang}, {Watanabe},
  {Weinberg}, {Yanny}, {Yasuda}, \& {SDSS Collaboration}}]{york2000}
{York}, D.~G., \& {SDSS Collaboration}. 2000, \aj, 120, 1579

\bibitem[{{Zakamska} {et~al.}(2004){Zakamska}, {Strauss}, {Heckman},
  {Ivezi{\'c}}, \& {Krolik}}]{zakamska2004}
{Zakamska}, N.~L., {Strauss}, M.~A., {Heckman}, T.~M., {Ivezi{\'c}}, {\v Z}.,
  \& {Krolik}, J.~H. 2004, \aj, 128, 1002

\bibitem[{{Zakamska} {et~al.}(2003){Zakamska}, {Strauss}, {Krolik}, {Collinge},
  {Hall}, {Hao}, {Heckman}, {Ivezi{\'c}}, {Richards}, {Schlegel}, {Schneider},
  {Strateva}, {Vanden Berk}, {Anderson}, \& {Brinkmann}}]{zakamska2003}
{Zakamska}, N.~L., {Strauss}, M.~A., {Krolik}, J.~H., {Collinge}, M.~J.,
  {Hall}, P.~B., {Hao}, L., {Heckman}, T.~M., {Ivezi{\'c}}, {\v Z}.,
  {Richards}, G.~T., {Schlegel}, D.~J., {Schneider}, D.~P., {Strateva}, I.,
  {Vanden Berk}, D.~E., {Anderson}, S.~F., \& {Brinkmann}, J. 2003, \aj, 126,
  2125

\end{thebibliography}

% TABLES AND FIGURES --------------------------------------

% ------------------------------------------------------------------------

\end{document}